\documentclass[graphicx,amsmath,amssymb,reprint,onecolumn]{revtex4-1}

\usepackage{graphicx}
\usepackage{color}
\usepackage{wasysym}
\usepackage{bm}
\usepackage{dsfont }
\usepackage[hang,FIGTOPCAP]{subfigure}
\usepackage{ulem}

\newcommand{\be}{\begin{equation}}
\newcommand{\bea}{\begin{eqnarray}}
\newcommand{\eea}{\end{eqnarray}}
\newcommand{\ee}{\end{equation}}

\begin{document}
\title{Bistable protein distributions in rod-shaped bacteria}

\author{L. Wettmann}
\affiliation{Theoretische Physik, Universit\"at des Saarlandes, Postfach 151150,
66041 Saarbr\"ucken, Germany}

\author{M. Bonny}
\affiliation{Theoretische Physik, Universit\"at des Saarlandes, Postfach 151150,
66041 Saarbr\"ucken, Germany}

\author{K. Kruse}
\affiliation{Theoretische Physik, Universit\"at des Saarlandes, Postfach 151150,
66041 Saarbr\"ucken, Germany}

\date{\today}

\begin{abstract}
The distributions of many proteins in rod-shaped bacteria are far from homogenous. Often they accumulate at the
cell poles or in the cell center. At the same time, the copy number of proteins in a single cell is relatively small making 
the patterns noisy. To explore limits to protein patterns due to molecular noise, we studied a generic mechanism 
for spontaneous polar protein assemblies in rod-shaped bacteria, which is based on cooperative binding of proteins 
to the cytoplasmic membrane. For mono-polar assemblies, we find that the switching time between the two poles 
increases exponentially with the cell length and with the protein number.   
\end{abstract}

\pacs{??}

\maketitle

\section{Introduction}

Though usually small in size, bacteria display a remarkable degree of spatial order~\cite{Shapiro:2009bj,kruse12}. In the simplest case, 
proteins are in a controlled way unevenly distributed between the two halves of rod-shaped bacteria like \textit{Escherichia coli} or 
\textit{Bacillus subtilis}. Examples 
for asymmetric distributions are provided by the accumulation of chemotactic receptors at the poles of \textit{E. coli}~\cite{Maddock:1993tu}, 
the localisation of Spo0J/Soj proteins close to the poles of \textit{B. subtilis}~\cite{Quisel:1999uh,Marston:1999uj}, or the Min 
proteins in \textit{E.~coli}, which periodically switch between the two cell halves~\cite{Raskin:1999vr}. More refined structures are the 
Z-ring that localises at the site of cell division~\cite{Bi:1991fr}, the localisation as a string of pearls of magnetosomes involved in bacterial 
magnetotaxis~\cite{Bazylinski:2004jx}, and the distribution of flagella of various motile bacteria~\cite{Chevance:2008bq}.  

Typically, the spatial distribution of proteins is linked to their function. Tom Duke was among the first to realise the importance of  
spatial structures for signalling. In the context of bacterial chemotaxis, he argued that the formation of receptor complexes increases
sensitivity~\cite{Duke:1999td}. Another example is provided by the Min-protein oscillations that are used to select the site of cell 
division~\cite{Loose:2011dd}.  To give just one more example, the linear arrangement 
of magnetosomes produces a strong enough magnetic dipole moment for bacteria to orient along the field lines of the geomagnetic 
field~\cite{Bazylinski:2004jx}. 

As a consequence of their functional importance, these spatial structures need to be maintained for a 
proper working of the cell. However, usually, the relatively small copy number of proteins in a cell, which typically ranges between less 
than ten to a few thousands, implies that molecular noise may severely restrict their life-time. In some cases, noise does not seem to 
be a problem. For example, the Spo0J/Soj proteins stochastically switch between the two nucleoids without impairing the fitness of 
\textit{B.~subtilis}~\cite{Quisel:1999uh,Marston:1999uj}. In other cases, molecular mechanisms assure the stability of protein patterns.
For instance, numerical analysis of the Min-protein dynamics has revealed that the oscillatory pattern is robust against molecular 
noise~\cite{Kerr:2006bh,Fange:2006ii,Arjunan:2010fd}. So, we face the question of how noise affects spatial protein distributions.

The influence of molecular noise on bacterial processes has intensively been studied during recent years~\cite{Eldar:2010kk}. 
In this context, a particular focus has been put on bistable systems~\cite{Balazsi:2011bw}. Spatially extended systems 
have received less attention, though. One example is provided by the Spo0J/Soj system mentioned 
above~\cite{Quisel:1999uh,Marston:1999uj,Doubrovinski:2005cm}. Also the Min proteins in \textit{E.~coli} were studied in this context
and found to stochastically switch between the two cell halves in sufficiently small cells after moderate 
over-expression~\cite{FischerFriedrich:2010hv,Sliusarenko:2011cz,Bonny:2013gt}.

The spatial structures mentioned above have in common that the proteins in question assemble on a support, for example, the 
membrane or the nucleoid. Consequently, spatial cues on these scaffolds might underlie the formation of the protein aggregates. In 
an extreme case, the proteins would not interact with each other but rather move in a potential landscape imposed by the spatial cues 
on the scaffold. An alternative is protein self-organisation~\cite{kruse12}. Mechanisms for self-organisation have been 
proposed for Spo0J/Soj~\cite{Doubrovinski:2005cm} and the Min proteins~\cite{Howard:2005dd}. For the latter,
the possibility of self-organisation has been demonstrated in reconstitution experiments~\cite{Loose:2008ca,Zieske:2013fi}.

In this work, we study the switching between two self-organised states of a spatially extended system in the weak-noise limit. In this case, 
typically, one cannot use Kramers rate theory, which relies on the existence of a potential. Instead, a generalisation of Kramers'
theory based on a pseudo-potential method can be employed~\cite{BENJACOB:1982vo,KUPFERMAN:1992wy,Maier:1993vv}. In a 
biological context, this 
method has been applied to bistable genetic switches~\cite{Roma:2005jj,Assaf:2011iw} and to bidirectional transport of molecular 
motors~\cite{Guerin:2011ed,Guerin:2011bk}. Our study is motivated by the heterogenous protein distributions in rod-shaped bacteria
described above. Heterogeneity of the distributions results from cooperative protein attachment to the membrane, which is motivated by 
studies of the Spo0J/Soj dynamics~\cite{Doubrovinski:2005cm} and the Min system~\cite{Huang:2003bc}. The molecular origin of cooperative 
binding remains to be understood, but has been very successful in these contexts. After defining the model, we will first perform stochastic 
simulations. To get further insight, we will perform a mean-field analysis and then follow the approach in~\cite{Hildebrand:1996ck} to establish 
the corresponding Fokker-Planck equation for the dominant modes.  Employing a WKB \textit{ansatz}, we will solve the Fokker-Planck equation 
and obtain the switching time as in Ref.~\cite{Maier:1993vv}. The work concludes with some remarks about possible generalisations.

\section{Stochastic dynamics for molecules forming membrane clusters}
\label{sec:model}
 
\subsection{The Chemical Master Equation}

\begin{figure}
\includegraphics[width=.5\linewidth]{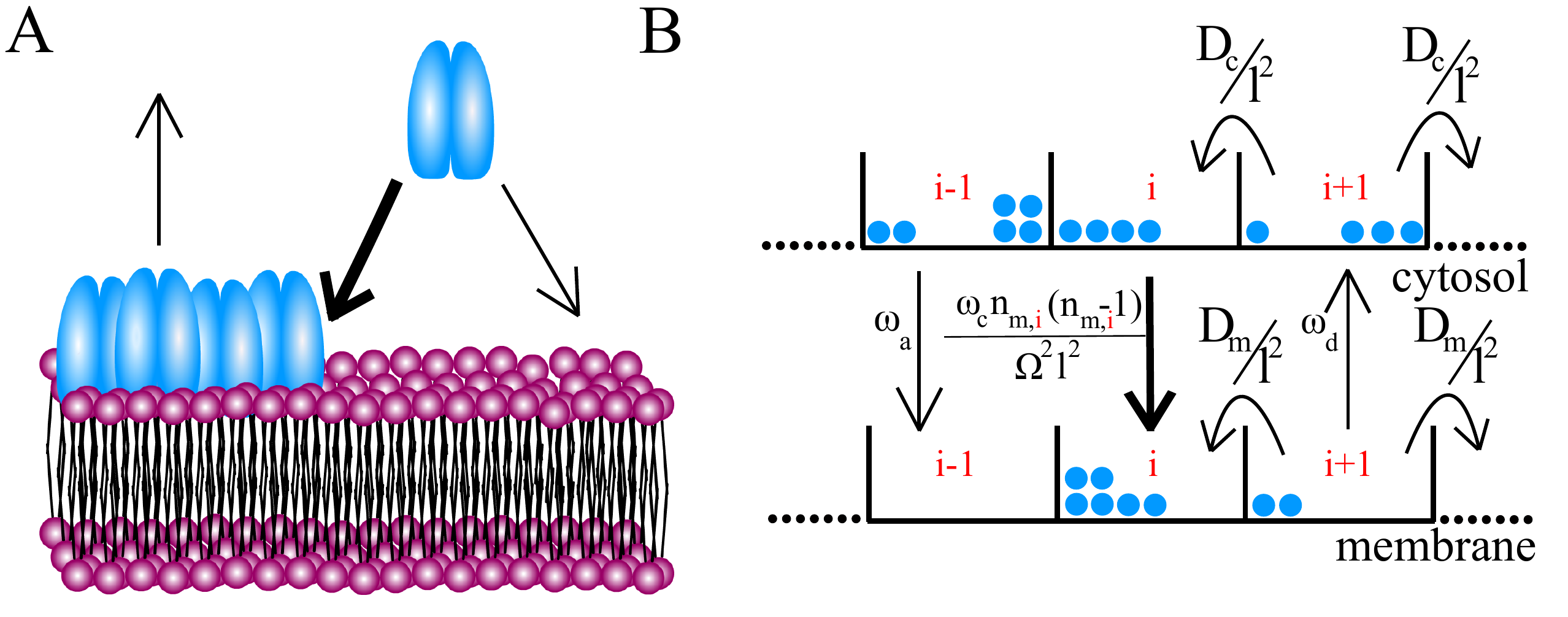}
\caption{Illustration of the system dynamics and geometry. A) Illustration of the molecular processes considered in this work.
They comprise spontaneous and cooperative attachment to as well as spontaneous detachment from the membrane.
B) Schematic of the geometry considered. The system consists of two linear arrays of bins of length $\ell$ that are assumed 
to be well mixed. The two arrays, respectively, represent the cytoplasm and the membrane. The rates for the different processes
are indicated.
\label{fig:sketch}}
\end{figure}

In the following, we consider the following processes: 1) Cytoplasmic molecules can bind to the membrane, with binding being 
favoured in regions, where membrane-bound molecules are already present; 2) spontaneous detachment of membrane-bound
molecules; 3) diffusion of cytoplasmic and membrane-bound molecules, see Fig.~\ref{fig:sketch}. We will consider the dynamics
in rod-like bacteria like \textit{E.~coli} and approximate the shape by a cylinder of length $L$ and radius $R$, where we assume the top 
and the bottom to be unavailable for binding. Furthermore, we will restrict attention to situations where the protein distribution is
invariant with respect to rotations around the cylinder axis. Possible consequences of these assumptions are discussed below. 
In direction of the long axis, the membrane is decomposed into compartments of length $\ell$. We will assume that particles within 
a compartment are well-mixed. This implies that we consider processes on time-scales that are larger or on the order of 
$\ell^2/D_\mathrm{m}$, where $D_\mathrm{m}$ is the diffusion constant of membrane-bound molecules. For cytoplasmic molecules, 
we assume a compartment that extends from the membrane to the cell center and again, we will consider it to be well-mixed. 
This implies that we are restricted to time-scales larger or of the order of $\tau\approx R^2/D_\mathrm{c}$, where 
$D_\mathrm{c}$ is the cytosolic diffusion constant. For $R\approx0.5\mu\mathrm{m}$ and $D_\mathrm{c}\approx15  
\mu\mathrm{m}^2/\mathrm{s}$ we get $\tau\approx20\mathrm{ms}$. 

The rate of spontaneous attachment to the membrane is denoted by $\omega_\mathrm{a}$, whereas the rate of spontaneous detachment 
is $\omega_\mathrm{d}$. The rate of co-operative membrane-binding in compartment $i$ is given by 
$\omega_\mathrm{c}n_{\mathrm{m},i}(n_{\mathrm{m},i}-1)/(\Omega^2\ell^2)$. In this expression, $n_{\mathrm{m},i}$ is the number of 
membrane-bound 
molecules in compartment $i$ -- the corresponding number of cytosolic molecules will be denoted by $n_{\mathrm{c},i}$ -- and $\Omega$ 
is the total number of molecules in the system. We make the scaling of the co-operative binding term with the compartment size and 
the total molecule number explicit, because we will later consider the limit of large molecule numbers $\Omega\to\infty$ and the continuum 
limit $\ell\to0$. 

The corresponding Chemical Master Equation for the evolution of the joint probability 
$P(\{n_{\mathrm{c},i}\}_{i=1,\ldots,N},\{n_{\mathrm{m},i}\}_{i=1,\ldots,N};t)$ can be written as
\begin{align}
\frac{dP}{dt} &= \sum_i \omega_\mathrm{a} \big( \mathbb{E}_{\mathrm{c},i}^+  \mathbb{E}_{\mathrm{m},i}^- -1\big) n_{\mathrm{c},i} P  
		+ \sum_i \frac{\omega_\mathrm{c}}{\Omega^2 \ell^2} \big( \mathbb{E}_{\mathrm{c},i}^+  \mathbb{E}_{\mathrm{m},i}^- -1\big) 
		n_{\mathrm{c},i} n_{\mathrm{m},i} (n_{\mathrm{m},i}-1) P 
		+ \sum_i \omega_\mathrm{d} \big( \mathbb{E}_{\mathrm{c},i}^-  \mathbb{E}_{\mathrm{m},i}^+ -1\big) n_{\mathrm{m},i} P \nonumber\\
		&+ \sum_i \frac{D_\mathrm{c}}{\ell^2} \mathbb{E}_{\mathrm{c},i}^-  \mathbb{E}_{\mathrm{c},i+1}^+  n_{\mathrm{c},i+1}P 
		+ \sum_i \frac{D_\mathrm{c}}{\ell^2} \mathbb{E}_{\mathrm{c},i}^-  \mathbb{E}_{\mathrm{c},i-1}^+  n_{\mathrm{c},i-1}P 	
		-2 \sum_i \frac{D_\mathrm{c}}{\ell^2} n_{\mathrm{c},i}P \nonumber	\\
		&+ \sum_i \frac{D_\mathrm{m}}{\ell^2} \mathbb{E}_{\mathrm{m},i}^-  \mathbb{E}_{\mathrm{m},i+1}^+  n_{\mathrm{m},i+1}P 
		+ \sum_i \frac{D_\mathrm{m}}{\ell^2}  \mathbb{E}_{\mathrm{m},i}^-  \mathbb{E}_{\mathrm{m},i-1}^+  n_{\mathrm{m},i+1}P	
		-2 \sum_i \frac{D_\mathrm{m}}{\ell^2} n_{\mathrm{m},i}P	\label{eq:CME}
\end{align}
In this expression $\mathbb{E}_{\mathrm{c},i}^\pm$ denote the particle creation and annihilation operators in compartment $i$ 
such that, for example, 
$\mathbb{E}_{\mathrm{c},i}^+f(\{n_{\mathrm{c},j}\}_{j=1,\ldots,N},\{n_{\mathrm{m},j}\}_{j=1,\ldots,N}) = 
f(\ldots,n_{\mathrm{c},i}+1,\ldots,\{n_{\mathrm{m},j}\}_{j=1,\ldots,N})$.
The corresponding operators for membrane bound compartments are denoted by $\mathbb{E}_{\mathrm{m},i}^\pm$. The terms
in the first line of the Chemical Master Equation account for attachment to and detachment from the membrane, the next two lines 
capture diffusion of cytoplasmic and membrane-bound molecules. 

\subsection{Numerical solution of the chemical Master equation}

We numerically analyse the Chemical Master Equation by combining the Gillespie algorithm with the next subvolume
method~\cite{Elf:2004fv}. Explicitly, in a given state, we determine for each compartment $j$ the total rate of a change in the number of
either the bound or unbound particles. For each compartment, we then draw a random number to determine the time at
which the next event occurs and perform the corresponding action for the compartment with the shortest waiting time. We
then go back and update the total rates of all compartments and continue as before.

\begin{figure}
\includegraphics[width=.5\linewidth]{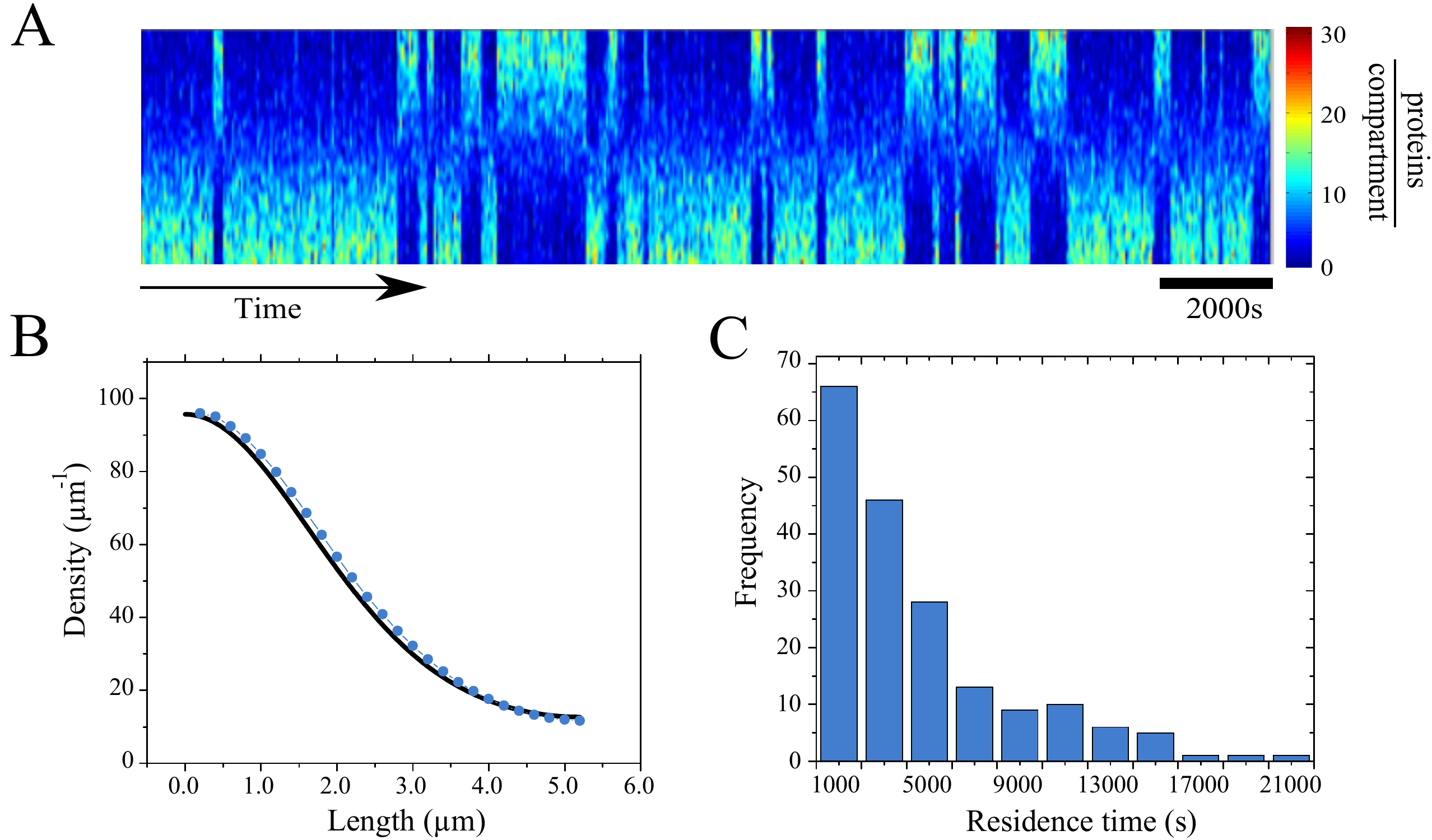}
\caption{Stochastic dynamics of membrane clustering. A) Colour-coded protein distribution as a function of time. B) 
Time-averaged profile of proteins residing in one cell half (blue dots) and mean field profile (black line). C) Distribution
of residence times. 
Parameters are $D_\mathrm{c}=10\mu$m$^2$s$^{-1}$, $D_\mathrm{c}=0.1\mu$m$^2$s$^{-1}$, $\omega_\mathrm{a}=0.1$s$^{-1}$, 
$\omega_\mathrm{d}=0.1$s$^{-1}$, $\omega_\mathrm{c}=1 \mu$m$^2$s$^{-1}$, $L=5.2\mu$m, $\ell=0.2\mu$m, $\Omega=187$ (A,C),
and $\Omega=260$ (B). \label{fig:stochastic}}
\end{figure}

In Figure~\ref{fig:stochastic} we present the results of a simulation for a system length of 5.2$\mu$m. The kymograph in 
Fig.~\ref{fig:stochastic}A shows that the particles accumulate in the vicinity of one end or ``cell pole". At random times, 
the particles switch to the opposite pole. The distribution of switching times has a mean value $\langle\tau\rangle=4965$s and is 
rather broad, see Fig.~\ref{fig:stochastic}C, with a standard deviation $\sigma_\tau=5487$s. The time-averaged profile of 
particles in one cell half agrees well with the profile obtained from the mean-field equations (\ref{eq:dcdtDim}) and (\ref{eq:dmdtDim}),
see Fig.~\ref{fig:stochastic}B, which will be introduced and discussed below. It should be noted that, as for all continuum theories, 
the agreement between
the mean-field theory and the stochastic simulations depends  on the choice of $\ell$ in the stochastic simulations. 
Each bin should contain on average a sufficiently large number of molecules such that a continuum approach is appropriate.
At the same time $\ell$ should be small compared to spatial features of the protein distribution. 

\section{The functional Fokker-Planck equation}

In order to go beyond the numerical analysis of the system, we will in this section derive a functional Fokker-Planck Equation 
from the Chemical Master Equation (\ref{eq:CME}). To this end will perform in a first step an expansion in terms of the inverse 
molecule number $1/\Omega$, which is a form of van Kampen's system size expansion. In a second step, we will take the continuum
limit $\ell\to0$. The resulting equation will first be analysed in the deterministic limit $\Omega\to\infty$. Then, we will use a Galerkin 
\textit{ansatz} to obtain an equation that is amenable to analysis of the influence of molecular noise on the system behaviour, which 
will be carried out in the next section. 

\subsection{System size expansion and continuum limit}

Let us consider the case of $\Omega\gg1$. In each compartment with large occupancies, $n_{\mathrm{c},i}\gg1$ and 
$n_{\mathrm{m},i}\gg1$, we can approximate the creation and annihilation operators of cytosolic and membrane-bound molecules, 
$\mathbb{E}_{\mathrm{c},i}^\pm$ and $\mathbb{E}_{\mathrm{m},i}^\pm$, respectively, by
\begin{align}
\mathbb{E}_{\mathrm{c},i}^\pm &\approx 1 \pm \frac{\partial}{\partial n_{\mathrm{c},i}} + 
\frac{1}{2} \frac{\partial^2}{\partial n_{\mathrm{c},i}^2} \\
\mathbb{E}_{\mathrm{m},i}^\pm &\approx 1 \pm \frac{\partial}{\partial n_{\mathrm{m},i}} + 
\frac{1}{2} \frac{\partial^2}{\partial n_{\mathrm{m},i}^2} \quad.
\end{align}
Formally, this corresponds to an expansion in the inverse molecule number up to second order. For large but finite values of 
$\Omega$, the above condition might not be fulfilled for all compartments at all times and higher orders of the approximation
should in principle be taken into account. Still, we will see that the ensuing equations give a quantitative account of the system 
behaviour. In the same spirit, we will also replace from now on the factor $n_{\mathrm{m},i}-1$ in the co-operative binding term 
by $n_{\mathrm{m},i}$. Explicitly, the Fokker-Planck Equation resulting from the Chemical Master Equation (\ref{eq:CME}) reads
\begin{align}
\frac{dP}{dt} = &- \sum_i \frac{\partial}{\partial n_{\mathrm{c},i}} \left\{D_\mathrm{c} \partial_x^2 n_{\mathrm{c},j} - 
			\omega_\mathrm{a} n_{\mathrm{c},i} + \frac{\omega_\mathrm{c}}{\Omega^2 \ell^2} n_{\mathrm{c},i} n_{\mathrm{m},i}^2 - 
			\omega_\mathrm{d} n_{\mathrm{m},i} \right\}P  \nonumber\\
			&- \sum_i \frac{\partial}{\partial n_{\mathrm{m},i}} \left\{D_\mathrm{m} \partial_x^2 n_{\mathrm{m},i}+
			\omega_\mathrm{a} n_{\mathrm{c},i} + \frac{\omega_\mathrm{c}}{\Omega^2 \ell^2} n_{\mathrm{c},i} n_{\mathrm{m},i}^2 - 
			\omega_\mathrm{d} n_{\mathrm{m},i} \right\}P  \nonumber\\
			&+ \sum_i \frac{\partial^2}{\partial n_{\mathrm{c},i}^2} \left\{D_\mathrm{c} \partial_x^2 n_{\mathrm{c},j} +
			\omega_\mathrm{a} n_{\mathrm{c},i} + \frac{\omega_\mathrm{c}}{\Omega^2 \ell^2} n_{\mathrm{c},i} n_{\mathrm{m},i}^2 + 
			\omega_\mathrm{d} n_{\mathrm{m},i} \right\}P  \nonumber\\
			&+ \sum_i \frac{\partial^2}{\partial n_{\mathrm{m},i}^2} \left\{D_\mathrm{m} \partial_x^2 n_{\mathrm{m},j} +
			\omega_\mathrm{a} n_{\mathrm{c},i} + \frac{\omega_\mathrm{c}}{\Omega^2 \ell^2} n_{\mathrm{c},i} n_{\mathrm{m},i}^2 + 
			\omega_\mathrm{d} n_{\mathrm{m},i} \right\}P  \nonumber\\
			&- \sum_i \frac{\partial^2}{\partial n_{\mathrm{c},i} \partial n_{\mathrm{m},i}} \left\{ 
			\omega_\mathrm{a} n_{\mathrm{c},i} + \frac{\omega_\mathrm{c}}{\Omega^2 \ell^2} n_{\mathrm{c},i} n_{\mathrm{m},i}^2 + 
			\omega_\mathrm{d} n_{\mathrm{m},i} \right\}P\label{eq:CMEapprox}
\end{align}

We will not use the Fokker-Planck Equation in this form, but rather go on by taking the continuum limit $\ell\to0$. To this end, we introduce 
the densities $c_{i}=n_{\mathrm{c},i}/(\Omega\ell)$ and $m_{i}=n_{\mathrm{m},i}/(\Omega\ell)$ for $i=1,\dots,N$. They give the 
fraction of all molecules that are cytosolic or membrane-bound in compartment $i$, such that $\sum_{i=1}^N \left(c_i+m_i\right)\ell=1$. 
Note, that $c_i$ like $m_i$ is a line-density, because we assume uniformity of the distribution in radial direction of the cell. 

In the limit $\ell\to0$, the densities can be replaced by continuous functions $c$ and $m$ with $c(i\ell)=c_i$ and $m(i\ell)=m_i$. 
Correspondingly, sums in Eq.~(\ref{eq:CMEapprox}) are replaced by integrals, $\sum_{i=1}^N\ell f_i\to\int_0^L dx f(x)$,
and partial derivatives by functional derivatives according to
\begin{align}
\frac{\partial f}{\partial n_{\mathrm{c},i}} &= \frac{1}{\Omega\ell}\frac{\partial f}{\partial c_i} \to\frac{1}{\Omega}\frac{\delta f}{\delta c(i\ell)} \\
\frac{\partial^2 f}{\partial n_{\mathrm{c},i}\partial n_{\mathrm{c},i}} &= \frac{1}{\Omega^2\ell^2}\frac{\partial^2 f}{\partial c_i\partial c_j} 
\to\frac{1}{\Omega^2}\frac{\delta^2 f}{\delta c(i\ell)\delta c(j\ell)}
\end{align}
and analogously for the other partial derivatives.

The resulting functional Fokker-Planck Equation for the probability functional $P[c(x),m(x),t]$ reads:
\begin{align}
\frac{\partial P}{\partial t} = & -\int_0^L dx \left\{\frac{\delta}{\delta c(x)} A_\mathrm{c} + \frac{\delta}{\delta m(x)}A_\mathrm{m}\right\}P
\label{eq:functionalFPE}\\
& + \frac{1}{2\Omega}\int_0^L dx\int_0^L dy\left\{\frac{\delta^2}{\delta c(x)\delta c(y)} B_\mathrm{cc}
+2\frac{\delta^2}{\delta c(x)\delta m(y)} B_\mathrm{cm}
+\frac{\delta^2}{\delta m(x)\delta m(y)} B_\mathrm{mm}\right\}P\nonumber
\end{align}
with
\begin{align}
A_\mathrm{c} &= D_\mathrm{c} \partial_x^2 c(x)+\omega   \\
A_\mathrm{m} &= D_\mathrm{m} \partial_x^2 m(x)-\omega\\
B_\mathrm{cc} &= \left\{D_\mathrm{c}\partial_x\partial_{y}c(x)+\bar\omega\right\}\delta(x-y)\\
B_\mathrm{mm} & = \left\{D_\mathrm{m}\partial_x\partial_{y}m(x)+\bar\omega\right\}\delta(x-y)\\
B_\mathrm{cm}&=-2\bar\omega\delta(x-y),
\intertext{where}
\omega &= \omega_\mathrm{a} c + \omega_\mathrm{c} c m^2 - \omega_\mathrm{d} m\\
\bar\omega &=\omega_\mathrm{a} c + \omega_\mathrm{c} c m^2 + \omega_\mathrm{d} m.
\end{align}
The terms containing $A_\mathrm{c}$ and $A_\mathrm{m}$ describe the ``convective" or deterministic part, whereas the ``diffusion" 
terms containing $B_\mathrm{cc}$, $B_\mathrm{cm}$, $B_\mathrm{mc}$, and $B_\mathrm{mm}$ account for the influence of noise. 
Note, that the latter are proportional to the inverse total molecule number $1/\Omega$. In the limit $\Omega\to\infty$, we get 
$P[c(x),m(x),t]=P_0[c(x,t),m(x,t)]$, where $c(x,t)$ and $m(x,t)$ obey the time evolution equations in the deterministic limit 
$\partial_t c = -A_\mathrm{c}$ and $\partial_t m = -A_\mathrm{m}$, and where $P_0[c(x),m(x)]=P[c(x),m(x),t=0]$. 

\subsection{The deterministic limit}

We start the analysis of the functional Fokker-Planck equation by considering the deterministic limit. Explicitly, the dynamic equations
for the particle densities $c$ and $m$ read
\begin{align}
\partial_t c &= D_\mathrm{c} \partial^2_x c -(\omega_\mathrm{a} +\omega_\mathrm{c} m^2)c  + \omega_\mathrm{d} m   
\label{eq:dcdtDim}\\
\partial_t m &= D_\mathrm{m} \partial^2_x m +(\omega_\mathrm{a} +\omega_\mathrm{c} m^2)c  - \omega_\mathrm{d} m \quad.
\label{eq:dmdtDim}
\end{align}
They are complemented by no-flux boundary conditions on the diffusion currents $-D_\mathrm{c}\partial_x c=0$ and
$-D_\mathrm{m}\partial_x m=0$ at $x=0$ and $x=L$.

Introducing the dimensionless time $t'=t\omega_\mathrm{a}$ and space $x'=x/\lambda$ with $\lambda^2=D_\mathrm{c}/\omega_\mathrm{a}$
and dimensionless densities $c'$ and $m'$ through $c = c'/\sqrt{\omega_\mathrm{c}/\omega_\mathrm{a}}$ and $m = m'/\sqrt{\omega_\mathrm{c}/\omega_\mathrm{a}}$, 
the dynamic equations can be written in the dimensionless form
\begin{align}
\partial_t c & = \phantom{D}\partial^2_xc- (1 + m^2)c + k m 
\label{eq:dcdt}  \\
\partial_t m &= D\partial^2_xm+ (1 + m^2)c - k m\quad,
\label{eq:dmdt}
\end{align}
where we have dropped the primes for the ease of notation. The dimensionless diffusion constant is $D=D_\mathrm{m}/D_\mathrm{c}$
and the dimensionless detachment rate $k=\omega_\mathrm{d}/\omega_\mathrm{a}$. Note, that also the system size $L$ is now
measured in units of $\lambda$.

For a total density $C_\mathrm{tot}=\frac{1}{L}\int_0^L dx (c+m)$, the deterministic equations have a stationary homogenous 
solution $c=C_0=const$ and $m=M_0=const$ that are determined by $C_0=C_\mathrm{tot}-M_0$ and
\begin{align}
M_0^3 - C_\mathrm{tot}M_0^2+(1+k)M_0+C_\mathrm{tot}&=0.
\end{align}
For the parameter values used in this work, there is only one real solution to this equation.

For a linear stability analysis of the homogenous stationary solution, we expand the densities
\begin{align}
c(x,t) &=\sum_{n=0}^\infty C_n(t)\cos(n\pi x/L)\label{eq:Cexpand}\\
m(x,t) &=\sum_{n=0}^\infty M_n(t)\cos(n\pi x/L)\label{eq:Mexpand}
\end{align}
and keep only linear terms in $C_n$ and $M_n$, $n=1,2,\ldots$ in the dynamic equations
\begin{align}
\frac{d}{dt}\left(\begin{array}{c} C_n \\ M_n\end{array}\right) &=
\left(\begin{array}{cc} -q_n^2-1-M_0^2 & -2C_0M_0+k\\ 1+M_0^2 &-Dq_n^2+2C_0M_0-k\end{array}\right)
\left(\begin{array}{c} C_n \\ M_n\end{array}\right) ,
\end{align}
where $q_n=n\pi/L$ and $n=1,2,\ldots$ 

\begin{figure}
\includegraphics[width=.5\linewidth]{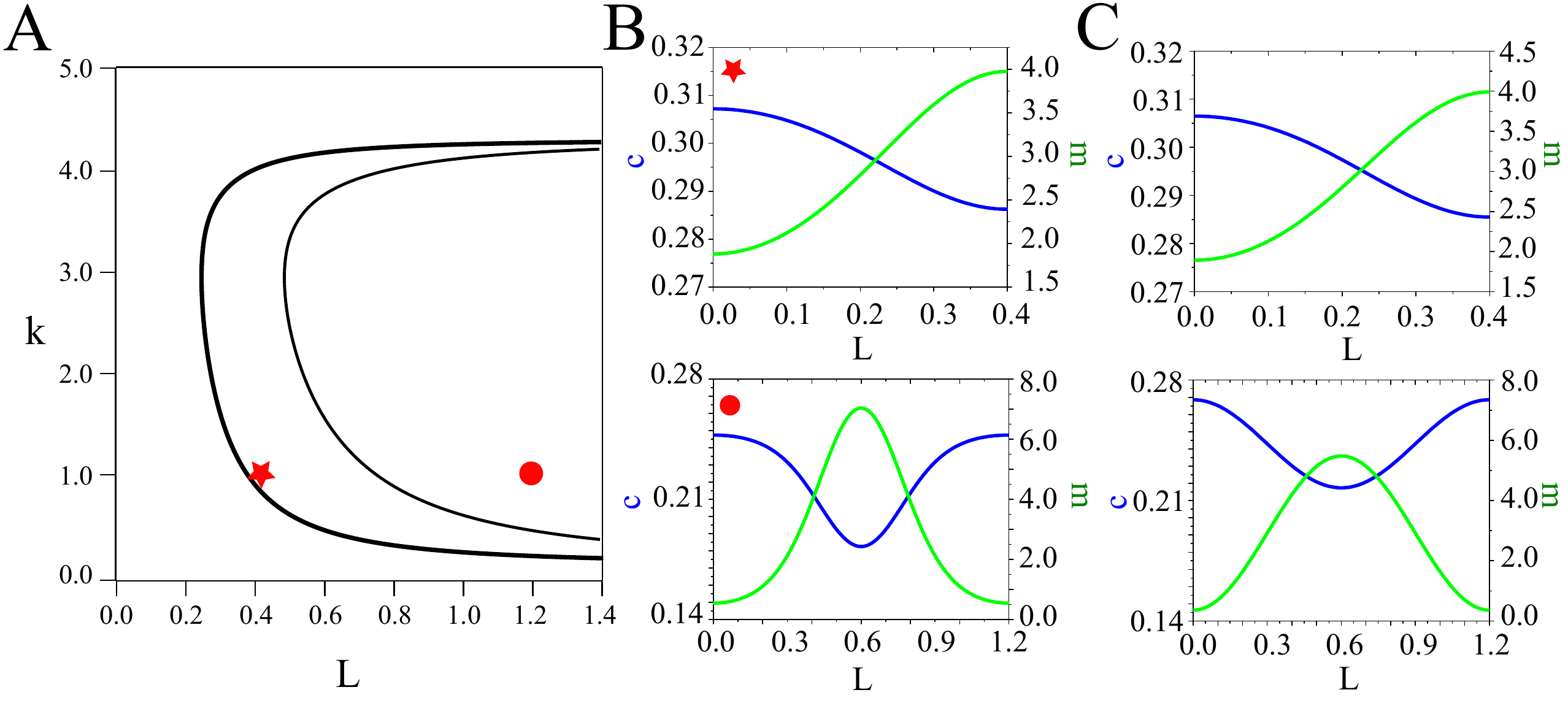}
\caption{System behaviour in the deterministic limit. A) State diagram as a function of the dimensionless system length $L$ and the dimensionless
detachment rate $k$. B) Steady states of the dynamic equations (\ref{eq:dcdt}) and (\ref{eq:dmdt}). C) Steady states of the reduced
equations (\ref{eq:dC0dt})-(\ref{eq:dM2dt}) where $C_0$, $C_1$, $C_2$, and $M_0$ have been adiabatically eliminated. Parameter
values for $k$ and $L$ in (B) and (C) are indicated by the symbols in (A), $D=0.01$, and $C_\mathrm{tot}=\sqrt{10}$ .
\label{fig:deterministic}}
\end{figure}

The homogenous state is stable for $k_{c,1}\ge k\ge k_{c,2}$ with some critical values $k_{c,1}$ and $k_{c,2}$. In between these 
two values, there is a critical system length $L_\mathrm{c}(k)$ beyond which 
the homogenous distribution becomes unstable through a pitchfork bifurcation, see Fig.~\ref{fig:deterministic}A. The first mode to become
unstable is the one with $n=1$ corresponding to situations in which the molecules accumulate at a pole. As the 
system is invariant under the transformation $x\to L-x$, there co-exist two mirror-symmetric solutions. 
One of them is chosen by spontaneous symmetry breaking, see Fig.~\ref{fig:deterministic}B.

Increasing the system length beyond $L_\mathrm{c}$, there is a second bifurcation at $L=L_{c,2}(k)$ through which two new solutions appear. 
They correspond to states, where the proteins either pile up in the cell center or accumulate at both poles. Numerical integration of 
the deterministic equations (\ref{eq:dcdt}) and (\ref{eq:dmdt}) indicates that only the state with proteins clustering in the center is stable, 
see Fig.~\ref{fig:deterministic}B.

We can make a Galerkin \textit{ansatz} and truncate the series (\ref{eq:Cexpand}) and (\ref{eq:Mexpand}) at a finite value of $n$.
The dynamic equations resulting after insertion of the truncated series in Eqs.~(\ref{eq:dcdt}) and (\ref{eq:dmdt}) are given in 
Appendix~\ref{app:Galerkin}. It turns out that for $L\lesssim L_{c,2}$ a series with $n=2$ gives a faithful description of the system 
dynamics, see Fig.~\ref{fig:deterministic}C. Furthermore, the modes $C_0$, $C_1$, $C_2$, and $M_0$ relax on faster time scales 
than the modes $M_1$ and $M_2$, such that we can make an adiabatic approximation and express the former in terms of the latter. 
The dynamic equations thus only depend on $M_1$ and $M_2$. 

\begin{figure}
\includegraphics[width=.75\linewidth]{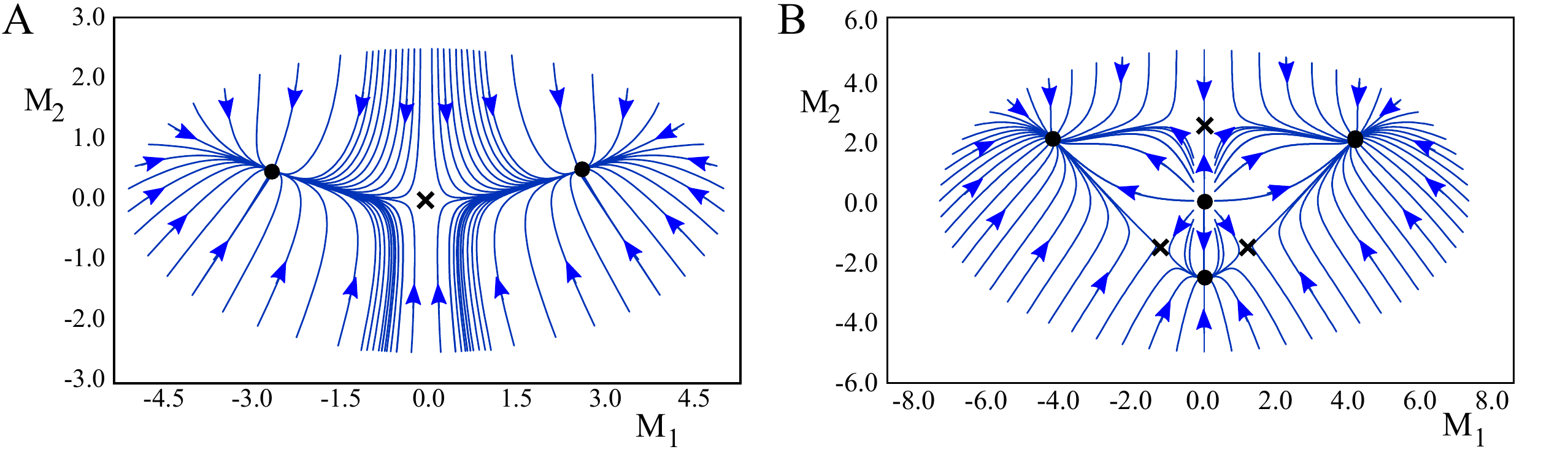}
\caption{Flow field for the modes $M_1$ and $M_2$ in the deterministic limit for $L_\mathrm{c}\le L\le L_{c,2}$ (A) and $L_{c,2}\le L$ (B).
Black dots indicate stable and unstable fixed points, black crosses hyperbolic fixed points. Parameters are as in Fig.~\ref{fig:deterministic} 
and $L=0.52$ (A) and $L=1.2$ (B).
\label{fig:flowfield}}
\end{figure}

In Figure~\ref{fig:flowfield}, we present the flow field for the dynamic equations for $M_1$ and $M_2$ after adiabatic elimination
of $C_0$, $C_1$, $C_2$, and $M_0$. For $L_\mathrm{c}\le L\le L_{c,2}$, the system has two symmetric stable and one hyperbolic fixed point,
see Fig.~\ref{fig:flowfield}A. 
The hyperbolic fixed point is at $(0,0,0,0)$ and corresponds to the homogenous state. The symmetric fixed points have $|M_2|\ll |M_1|$
such that they correspond to protein accumulations at either of the two poles. As soon as $L>L_{c,2}$ two new fixed points appear.
These hyperbolic fixed points have $M_1=0$ and correspond to symmetric protein distribution with, respectively, accumulation in the center
and simultaneous accumulation at both poles. Further increase of $L$ leads to two new hyperbolic fixed points. At the same time,
the fixed point with $M_1<0$ becomes stable. It should be noted, though, that the corresponding symmetric distribution is not observed in
the stochastic simulations, indicating that this fixed point remains hyperbolic if higher modes are taken into account. The two new hyperbolic
fixed points are connected to the stable fixed points by heteroclinic orbits, see Fig.~\ref{fig:flowfield}B.

\subsection{Galerkin approximation for the Fokker-Planck equation}

We will now exploit the findings of the previous subsection and use the Galerkin approximation for the Fokker-Planck 
Equation~(\ref{eq:functionalFPE}). We will use the same dimensionless parameters 
and densities as above. Employing Equations (\ref{eq:Cexpand}) and (\ref{eq:Mexpand}), we can transform the functional $P[c(x),m(x);t]$
into a function of the coefficients $C_n$ and $M_n$, $n=0,1,2,\ldots$ The corresponding Fokker-Planck Equation can be derived
from Eq.~(\ref{eq:functionalFPE}). To this end, we first express the functional derivatives with respect to $c(x)$ and $m(x)$ in terms
of partial derivatives with respect to $C_n$ and $M_n$, $n=0,1,2,\ldots$ Considering a variation $\delta C_n$ of mode $n$ of 
membrane-bound molecules yields $\delta c(x)=\delta C_n\cos(n\pi x/L)$. If $\bar F$ denotes the function of $C_n$ and $M_n$ 
corresponding to a functional $F[c(x),m(x)]$, then
\begin{align}
\frac{\partial \bar F}{\partial C_n} &=\int_0^L dx \frac{\delta F}{\delta c(x)}\cos\left(\frac{n\pi x}{L}\right)
\end{align}
and analogously for the derivatives $\partial \bar F/\partial M_n$. Note, that variations of $C_0$ and $M_0$ are not independent
as $C_0+M_0=1/L$. From the above expression follows that
\begin{align}
\frac{\delta F}{\delta c(x)} &= \frac{2}{L}\sum_{n=0}^\infty \frac{\partial \bar F}{\partial C_n}\cos\left(\frac{n\pi x}{L}\right)\\
\frac{\delta F}{\delta m(x)} &= \frac{2}{L}\sum_{n=0}^\infty \frac{\partial \bar F}{\partial M_n}\cos\left(\frac{n\pi x}{L}\right)
\intertext{and similarly}
\frac{\delta^2 F}{\delta c(x)\delta c(y)} &= \frac{4}{L^2}\sum_{n,m=0}^\infty \frac{\partial^2 \bar F}{\partial C_n\partial C_m}
\cos\left(\frac{n\pi x}{L}\right)\cos\left(\frac{m\pi x}{L}\right)\\
\frac{\delta^2 F}{\delta c(x)\delta m(y)} &= \frac{4}{L^2}\sum_{n,m=0}^\infty \frac{\partial^2 \bar F}{\partial C_n\partial M_m}
\cos\left(\frac{n\pi x}{L}\right)\cos\left(\frac{m\pi x}{L}\right)\\
\frac{\delta^2 F}{\delta m(x)\delta m(y)} &= \frac{4}{L^2}\sum_{n,m=0}^\infty \frac{\partial^2 \bar F}{\partial M_n\partial M_m}
\cos\left(\frac{n\pi x}{L}\right)\cos\left(\frac{m\pi x}{L}\right).
\end{align}

Inserting the expansions (\ref{eq:Cexpand}) and (\ref{eq:Mexpand}) into Eq.~(\ref{eq:functionalFPE}), using the expressions 
for the functional derivatives just derived, and performing the integrals, we obtain a Fokker-Planck Equation for the probability
density $P(\{C_n\}_{n=0,1,\ldots},\{M_n\}_{n=0,1,\ldots};t)$, where again $C_0+M_0=1/L$. It is of the form
\begin{align}
\frac{\partial P}{\partial t} =& 
-\sum_{n=0}^\infty \left\{\frac{\partial}{\partial C_n} \bar A_{\mathrm{c},n} + \frac{\partial}{\partial M_n}\bar A_{\mathrm{m},n}\right\}P
\label{eq:FPE}\\
&+ \frac{1}{2\Omega}\sum_{n,m=0}^\infty\left\{\frac{\partial^2}{\partial C_n\partial C_m} \bar B_{\mathrm{cc},nm}
+2\frac{\partial^2}{\partial C_n\partial M_m} \bar B_{\mathrm{cm},nm}
+\frac{\partial^2}{\partial M_n\partial M_m} \bar B_{\mathrm{mm},nm}\right\}P.\nonumber
\end{align}
We will now apply the Galerkin approximation and truncate the series at $n,m=2$. Furthermore, we will make the adiabatic 
approximation and express $C_0$, $C_1$, $C_2$, and $M_0$ in terms of $M_1$ and $M_2$ by using the deterministic 
equations (\ref{eq:dC0dt})-(\ref{eq:dM0dt}) in steady state. Eventually, we arrive at
\begin{align}
\label{eq:FPE12}
\partial_t P&=-\partial_k u_kP+\frac{1}{2\Omega L'}\partial_k\partial_l D_{kl}P,
\end{align}
where the probability distribution $P\equiv P(M_1,M_2;t)$ now depends only on $M_1$ and $M_2$ and where 
$\partial_k\equiv\partial/\partial M_k$ with $k=1,2$. Furthermore, $L'=L\sqrt{D_\mathrm{c}/\omega_\mathrm{c}}$ and summation over repeating 
indices is understood. The expressions for the drift velocity $u_k=\dot M_k$ are those given 
above in the deterministic limit, Eqs.~(\ref{eq:dM1dt}) and (\ref{eq:dM2dt}). The expressions for the diffusion matrix $\mathsf{D}$ are 
given in Appendix~\ref{app:Galerkin}. Note, that the elements $D_{kl}$ depend on $M_1$ and $M_2$.

In the next section, we will use the Fokker-Planck Equation (\ref{eq:FPE12}) to determine the average switching time.

\section{Estimation of the switching time in the weak noise limit}

The switching time can equivalently be interpreted as the mean first-passage time (MFPT) of a multi-dimensional escape problem.
However, in contrast to most cases studied in the literature, the deterministic dynamics is in the present case not determined by
a potential, so that we cannot use Kramers (or Eyring) rate theory. Instead, we will employ a generalised framework developed by
Maier and Stein~\cite{Maier:1993vv}. In a first step, we will present the necessary expressions and then compare their solution to data 
obtained from stochastic simulations. We continue to use the dimensionless parameters introduced above.

\subsection{Probability flux across the separatrix}

As we had seen above, in the deterministic case there is a region in parameter space, where two stable mirror-symmetric 
distributions co-exist. The two basins of attraction are separated by a separatrix along $M_1=0$, which contains another 
fixed point, namely the homogenous state $M_1=M_2=0$. It is a hyperbolic fixed point for which the stable manifold coincides 
with the separatrix. The aim is now to calculate the MFPT $\tau$ in the limit of weak noise. 

To this end, we consider the Fokker-Planck equation (\ref{eq:FPE12}) with absorbing boundary conditions along the separatrix. 
We will calculate the total probability current across the separatrix associated with the eigenfunction $P_1$ of the Fokker-Planck 
operator that decays most slowly. Normalising this current to the total probability to find a particle on one side of the separatrix 
yields
\begin{equation}
\tau = \frac{-\int_{-\infty}^\infty\frac{1}{2\Omega L'}\partial_kD_{1k}P(0,M_2) dM_2}{\int_0^\infty\int_{-\infty}^\infty P(M_1,M_2) dM_2 dM_1},
\label{eq:tau}
\end{equation}
where summation with respect to the index $k$ is understood. Note, that $\tau$ is just the characteristic time on which the eigenfunction
$P_1$ relaxes, that is, it is the inverse of the corresponding eigenvalue. In lowest order in $1/\Omega$, the 
eigenfunction $P_1$ can be replaced by the solution $P_\mathrm{s}$ to the steady-state Fokker-Planck Equation~\cite{Naeh:1990up}.

The steady state of Eq.~(\ref{eq:FPE12}) cannot be calculated exactly. Instead, we will make a Wentzel-Kramers-Brillouin ansatz
and write the steady state probability distribution as
\begin{align}
\label{eq:WKB}
P_\mathrm{s}(M_1,M_2) &= K(M_1,M_2)\exp\{-\Omega L' S(M_1,M_2)\}
\end{align}
with the classical action (or quasi-potential) $S$. The equation for the action is obtained by inserting the WKB ansatz into
Eq.~(\ref{eq:FPE12}) and by considering the terms of first order in $1/\Omega$. We get an equation of the form of a 
Hamilton-Jacobi Equation $H(M_1,M_2,\partial_1S,\partial_2S)=0$ with the Hamiltonian
\begin{align}
\label{eq:hamiltonian}
H&= u_kp_k+\frac{1}{2}D_{kl}p_kp_l,
\end{align}
where $p_k=\partial_kS$ is the momentum conjugated to $M_k$, $k=1,2$. The action is then obtained from solving first the
canonical dynamic equations, which in our case read
\begin{align}
\label{eq:dMdt}
\dot M_k &=u_k+D_{kl}p_l\\
\dot p_k &=-(\partial_ku_l)p_l-\frac{1}{2}(\partial_kD_{lm})p_lp_m,\\
\intertext{and then the dynamic equation}
\dot S&= D_{kl}p_kp_l+u_kp_k.
\end{align}

It remains to determine the pre-factor $K$ in Eq.~(\ref{eq:WKB}), which can be obtained from the terms of second order in $1/\Omega$
in Eq.~(\ref{eq:FPE12}). Explicitly,
\begin{align}
\label{eq:dKdt}
\dot K&=-\left[\partial_k (u_k+ D_{kl}p_l) + \frac{1}{2} W_{kl} D_{kl} \right] K.
\end{align}
In this expression $\mathsf{W}$ denotes the Hessian of the action $S$, that is, $W_{kl}=\partial_k\partial_lS$. It can be 
determined without knowledge of the action by solving its time evolution equation, which follows a (generalised) Riccati equation
\begin{align}
\label{eq:dWdt}
\dot{W_{kl}} = -W_{km} D_{mn}  W_{nl} - \left(\partial_k\dot M_n\right)W_{nl} 
 - \left(\partial_l \dot M_n\right) W_{nk} + \frac{1}{2}\left(\partial_l \dot p_k+\partial_k \dot p_l\right).
\end{align}

\begin{figure}
\includegraphics[width=.5\linewidth]{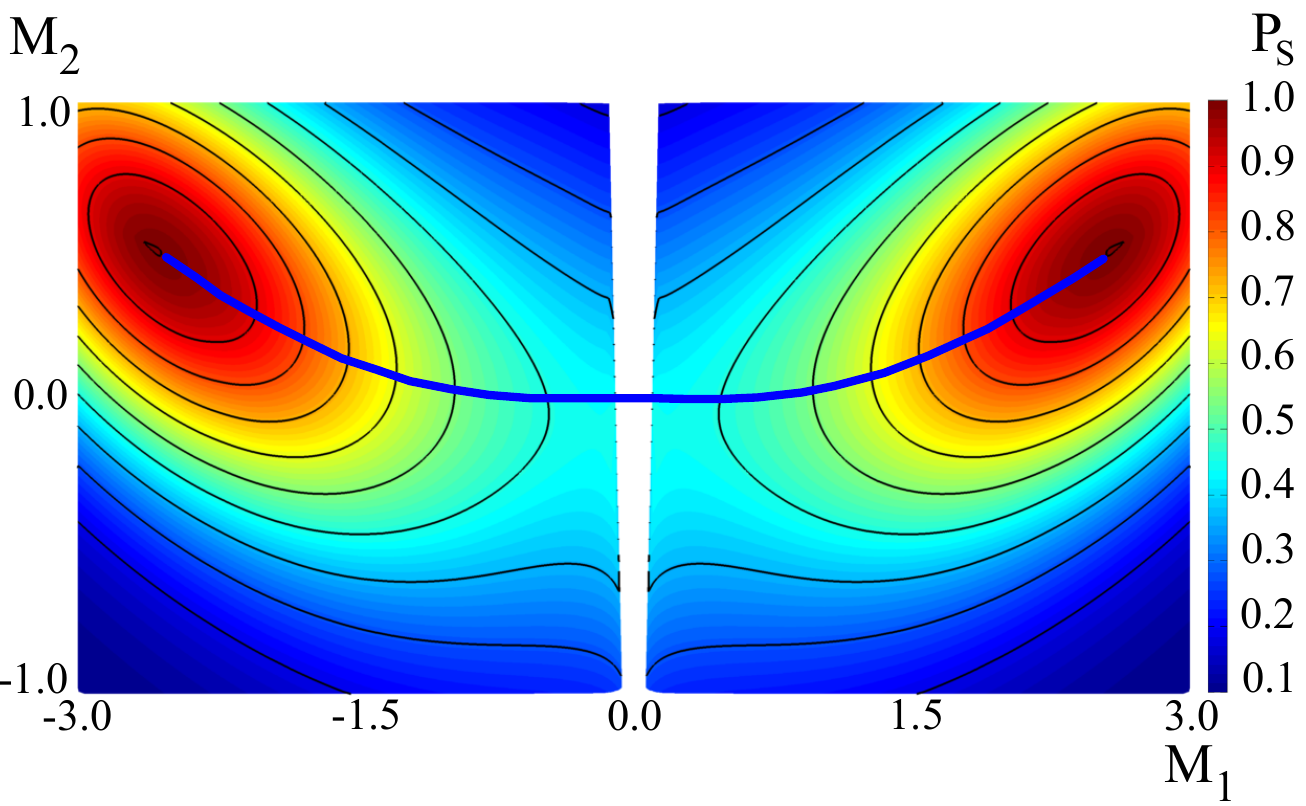}
\caption{Steady state probability distribution in the WKB approximation. The blue curve indicates the heteroclinic orbit
between the two stable fixed points passing through the hyperbolic fixed point at $(0,0,0,0)$. Parameters are as in Fig.~\ref{fig:deterministic}
with $L=0.52$ and $\Omega=30$.
\label{fig:probabilitydistribution}}
\end{figure}

In Figure~\ref{fig:probabilitydistribution}, we present the steady state probability distribution corresponding to the phase portraits 
displayed on Fig.~\ref{fig:flowfield}A. The probability is maximal in the vicinity of the stable fixed points. 
In the hyperbolic fixed point at $(0,0,0,0)$ the probability distribution has a saddle point (slight deviations are due to interpolation 
errors) and the heteroclinic orbit indicated in blue follows the ridge of probability distribution.

We can exploit the form of the probability distribution to further simplify Eq.~(\ref{eq:tau}). The integrand in the denominator of the 
above expression for $\tau$ is obviously dominated by the region around the stable fixed point, whereas the integrand in the numerator 
is dominated by the region around the hyperbolic fixed point $M_1=M_2=p_1=p_2=0$. In these regions, we can make a Gaussian 
approximation of the probability density or, equivalently, use a quadratic approximation of the Hamiltonian (\ref{eq:hamiltonian}): 
\begin{align}
\label{eq:appHamiltonian}
H &= \lambda_{kl} M_k p_l + D_{kl} p_k p_l,
\end{align}
where $\lambda_{kl}\equiv\partial_k u_l$ and where the coefficients are evaluated in the hyperbolic or stable fixed points of 
Eqs.~(\ref{eq:dM1dt}) and (\ref{eq:dM2dt}), respectively. With this approximation, the switching time is estimated as
\begin{align}
\label{eq:tauApp}
\tau &= \frac{\pi}{\lambda^\mathrm{hyp}_{11} K^\mathrm{hyp}} \sqrt{\frac{|\mathrm{det}\ \mathsf{W}^\mathrm{hyp}|}{\mathrm{det}\ 
\mathsf{W}^\mathrm{st}}} \mathrm{exp}\{\Omega L' S^\mathrm{hyp}\}.
\end{align}
In this expression, the quantities with a superscript 'hyp' are evaluated in the hyperbolic fixed point, whereas a superscript 'st' indicates 
evaluation in the stable fixed point. 

Finally, the expression for $\det\mathsf{W}^\mathrm{st}$ and $\det\mathsf{W}^\mathrm{hyp}$ can be obtained from 
Eq.~(\ref{eq:dWdt}) by setting the time-derivative equal to zero and using $p_1=p_2=0$. It yields
\begin{align}
W_{km} D_{mn} W_{nl} + \left(\partial_l u_n\right) W_{nk}+\left(\partial_k u_n\right)W_{nl}=0.
\end{align}
The elements $D_{mn}$ and the partial derivatives $\partial_m u_n$ again have to be evaluated at the respective fixed point.

We have now given all the elements necessary for estimating the switching time $\tau$.

\subsection{Comparison of the estimated switching time with stochastic simulations}

According to Eq.~(\ref{eq:tau}), we need to evaluate various quantities in the fixed points of Eqs.~(\ref{eq:dM1dt}) and (\ref{eq:dM2dt}).
In the hyperbolic fixed point, the approximate Hamiltonian (\ref{eq:appHamiltonian}) can be obtained analytically. 
The matrices $\mathsf{\lambda}$ and $\mathsf{D}$ are diagonal with non-zero elements
\begin{align}
\lambda^\mathrm{hyp}_{11} &= \frac{\pi^2}{L^2}\left(\frac{2 C_0 M_0 - k}{1+M_0^2 +  \frac{\pi^2}{L^2}}-D \right) \\
\lambda^\mathrm{hyp}_{22} &= \frac{4\pi^2}{L^2} \left(\frac{2 C_0 M_0 - k}{ 1+ M_0^2 + \frac{4\pi^2}{L^2}}-D\right) \\
D^\mathrm{hyp}_{11} = D^\mathrm{hyp}_{22} &= 4k M_0.
\end{align}
From this we get for the determinant of the Hessian
\begin{align}
\det \mathsf{W}^\mathrm{hyp} &=\frac{\lambda^\mathrm{hyp}_{11}\lambda^\mathrm{hyp}_{22}}{4k^2 M_0^2}
\end{align}
The corresponding expressions for the stable fixed point have to be calculated numerically. The values of $K^\mathrm{hyp}$ and 
$S^\mathrm{hyp}$ in Eq.~(\ref{eq:tauApp}) require integration of the dynamic equations (\ref{eq:dMdt})-(\ref{eq:dKdt}) along the 
heteroclinic orbit connecting the stable and the hyperbolic fixed points. Due to the non-linear character of the dynamic equations, 
these integrations also have to be performed numerically. 

To determine the heteroclinic orbit, we used initial conditions such that $M_1$ and $M_2$ were located in the stable fixed
point of Eqs.~(\ref{eq:dM1dt}) and (\ref{eq:dM2dt}) and the energy was fixed to zero. These conditions left one of the generalised
momenta $p_1$ and $p_2$ undetermined. Its value was obtained by minimising the distance of the corresponding trajectory
to the hyperbolic fixed point at the origin. 

\begin{figure}
\includegraphics[width=.75\linewidth]{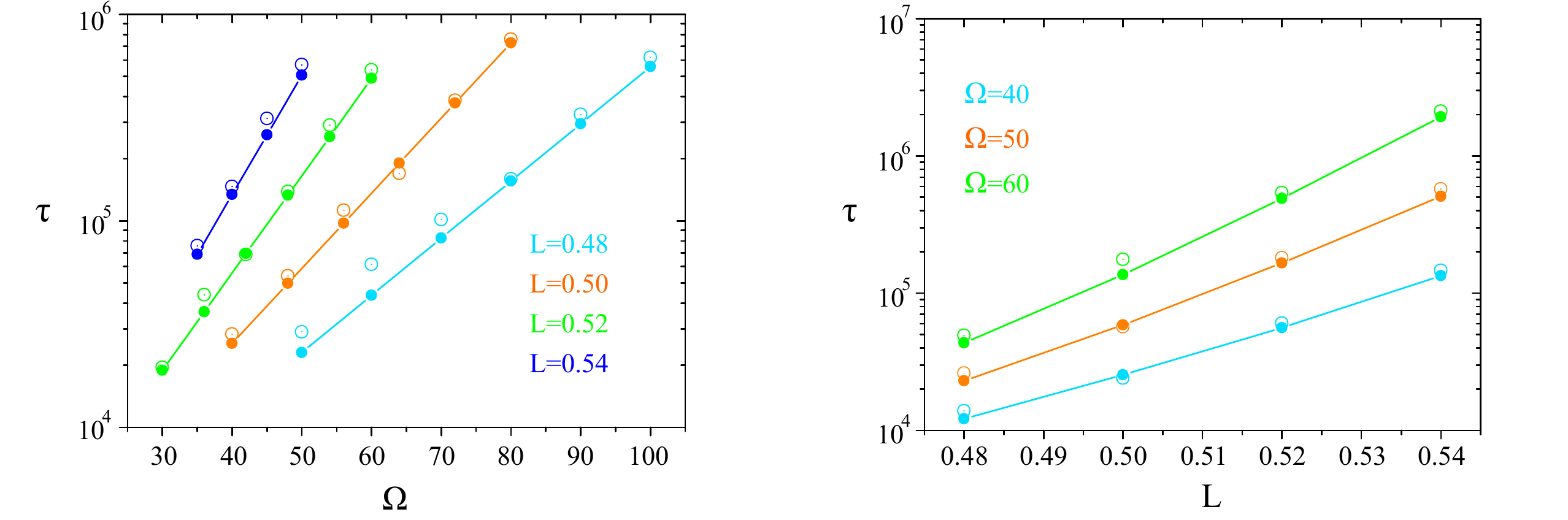}
\caption{\label{fig:comparison} Comparison between the estimated switching time (full circles) and switching times measured in 
stochastic simulations (open circles) as a function of the total number of molecules $\Omega$ (A) and the dimensionless system
length $L$ (B). }
\end{figure}

In Figure~\ref{fig:comparison}, we present the results of our calculation for various cell lengths. The agreement between the simulation
results and the approximate expression (\ref{eq:tau}) for the switching time is remarkably good taking into account the various
approximations we made along the way. The dependence of the switching time $\tau$ on the total molecule number $\Omega$ is
exponential as indicated by Eq.~(\ref{eq:tau}). Its dependence on the cell length $L$ is also essentially exponential, 
suggesting that its dominant contribution of $L$ to $\tau$ is given by the explicit dependence on $L$ in Eq.~(\ref{eq:tau}). This result
is in line with the observation made in Ref.~\cite{FischerFriedrich:2010hv}, that intracellular fluctuations decrease with increasing cell 
length.

\section{Discussion} 

In this work, we have studied the influence of molecular noise on the self-organised polar localisation of proteins in rod-shaped bacteria.
We found that cooperative attachment to the cell membrane is sufficient to generate asymmetric protein distributions along the 
bacterial long axis. Using stochastic simulations and an approximate analysis of the weak-noise limit we found that fluctuations in 
the protein distribution can lead to switches between two mirror-symmetric solutions. The frequency of these switches decreases
exponentially with the cell length. This finding is in line with experimental observations on the Min-protein system in 
\textit{E.~coli}~\cite{FischerFriedrich:2010hv}. It suggests that the stability of spatial patterns in rod-shaped bacteria depends more
on the total protein number in the cell than the protein density.

Here, we have restricted attention to the weak-noise limit, which is inherent to the use of a Fokker-Planck equation. It will be
interesting to see in future work, whether the approach can be extended to stronger noise. For example, one might use the
approach developed in Ref.~\cite{Wu:2013dx} to study the stability of the heterogenous states as a function of the total
protein number and estimate the minimal number of molecules needed to achieve polar localisation. 

The mechanism of self-organisation we have analysed, is based on cooperative binding of the molecules to a support and their 
spontaneous unbinding. Other molecular mechanisms might be considered. For example, we looked at the case that membrane-bound
proteins exist in two states. Proteins bind in a cooperative manner in one state to the membrane and then transit into a second state in which they
can unbind from the membrane. In a cellular context, the transition could be linked to the hydrolysis of ATP. If the transition 
between the two states occurs spontaneously, then the system behaviour is similar to the one discussed above. If proteins
in the second state can induce detachment of proteins in the first state, then the switching time will decrease with increasing
protein number and system length. This behaviour is similar to the one found for the Min proteins~\cite{FischerFriedrich:2010hv,Bonny:2013gt}.
Remarkably, if proteins in the second state only catalyse the transition of proteins in state one to state two, then the system
can spontaneously oscillate. Additional cooperative effects thus can have a significant influence on the system behaviour. 

A possible biological implication of our findings is that (self-organised) protein structures become less sensitive to
molecular noise as a cell ages. Indeed, as the cell grows, the number of molecules as well as the stability of 
molecular assemblies at one of the cell poles increase. Similarly Min-protein oscillations in \textit{E.~coli} were observed to become 
less noisy with increasing cell length~\cite{FischerFriedrich:2010hv}. It will be interesting to see, whether the techniques used
in the present work will allow us to gain a deeper understanding of this phenomenon, too.

\acknowledgments{Funding by Deutsche Forschungsgemeinschaft through the SFB 1027 is gratefully acknowledged.}

\appendix

\section{The dynamic equations in the Galerkin approximation}
\label{app:Galerkin}

As indicated in the main text, we can stop the expansion in Eqs.~(\ref{eq:Cexpand}) and (\ref{eq:Mexpand}) at $n=2$,
\begin{align}
c(x,t) &=C_0(t)+C_1(t)\cos(\pi x/L)+C_2(t)\cos(2\pi x/L)\\
m(x,t) &=M_0(t)+M_1(t)\cos(\pi x/L)+M_2(t)\cos(2\pi x/L),
\end{align}
and still capture the essential features of the system dynamics. Inserting these expressions in Eqs.~(\ref{eq:dcdt}) and 
(\ref{eq:dmdt}) and integrating with respect to $x$, we obtain
\begin{align}
\label{eq:dC0dt}
\dot C_0 &=-\left(1+M_0^2+\frac{1}{2}M_1^2+\frac{1}{2}M_2^2\right)C_0 -
\left(M_0M_1+\frac{1}{2}M_1M_2\right)C_1 -
\left(M_0M_2+\frac{1}{4}M_1^2\right) C_2+kM_0\\
\dot C_1 &= -\left(\frac{\pi}{L}\right)^2C_1 -\left(2M_0M_1+M_1M_2\right)C_0
-\left(1+M_0^2+M_0M_2+\frac{3}{4}M_1^2+\frac{1}{2}M_2^2\right)C_1
-\left(M_0M_1+M_1M_2\right)C_2 +kM_1\\
\dot C_2 &=-\left(\frac{2\pi}{L}\right)^2C_2-\left(2M_0M_2+\frac{1}{2}M_1^2\right)C_0 
-\left(M_0M_1+M_1M_2\right)C_1
-\left(1+M_0^2+\frac{1}{2}M_1^2+\frac{3}{4}M_2^2\right)C_2 +kM_2\\
\label{eq:dM0dt}
\dot M_0 &=\left(1+M_0^2+\frac{1}{2}M_1^2+\frac{1}{2}M_2^2\right)C_0 +
\left(M_0M_1+\frac{1}{2}M_1M_2\right)C_1 +
\left(M_0M_2+\frac{1}{4}M_1^2\right) C_2-kM_0\\
\label{eq:dM1dt}
\dot M_1 &= -D\left(\frac{\pi}{L}\right)^2M_1 +\left(2M_0M_1+M_1M_2\right)C_0
+\left(1+M_0^2+M_0M_2+\frac{3}{4}M_1^2+\frac{1}{2}M_2^2\right)C_1
+\left(M_0M_1+M_1M_2\right)C_2 -kM_1\\
\label{eq:dM2dt}
\dot M_2 &=-D\left(\frac{2\pi}{L}\right)^2M_2+\left(2M_0M_2+\frac{1}{2}M_1^2\right)C_0 
+\left(M_0M_1+M_1M_2\right)C_1
+\left(1+M_0^2+\frac{1}{2}M_1^2+\frac{3}{4}M_2^2\right)C_2 -kM_2
\end{align}
The coefficients of the diffusion matrix $\mathsf{D}$ in Eq.~(\ref{eq:FPE12}) then read
\begin{align}
D_{11} =& \left(2+2 M_0^2+ \frac{3}{2} M_1^2+ M_2^2+ 2 M_0 M_2\right)C_0 + \left(3 M_0 M_1+2 M_1 M_2 \right)C_1 \nonumber \\
&+ \left(1+  M_0^2+ M_1^2+ \frac{3}{4} M_2^2 + 2 M_0 M_2\right)C_2 + 2 k M_0 + \left(k  - 4\frac{\pi^2}{L^2} D \right)M_2 \\
D_{12} =D_{21}=& \left(2  M_0 M_1 +2 M_1 M_2\right)C_0+  \left(1+  M_0^2+ M_1^2+\frac{3}{4} M_2^2+ 2 M_0 M_2\right)C_1 \nonumber\\
&+ \left( 2 M_0 M_1+\frac{3}{2} M_1 M_2\right)C_2+ \left(k  - 4\frac{\pi^2}{L^2} D \right)M_1 \\
D_{22} =& \left(2+ 2 M_0^2+ M_1^2+ \frac{3}{2} M_2^2\right) C_0 + \left(2 M_0 M_1+ \frac{3}{2}  M_1 M_2\right)C_1\nonumber\\
&+\left(\frac{3}{4} M_1^2 + 3  M_0 M_2\right)C_2 + 2 k M_0,
\end{align}
where $M_0$, $C_0$, $C_1$, and $C_2$ are expressed in terms of $M_1$ and $M_2$ by solving Eqs.~(\ref{eq:dC0dt})-(\ref{eq:dM0dt})
in steady state.

\bibliographystyle{vancouver}
\bibliography{references.bib}
\end{document}